\documentclass{aa}  
\usepackage{graphicx}
\begin{document}

\title{Accretion-powered chromospheres in classical T Tauri stars
    \thanks{Based on observations collected at the Nordic Optical Telescope, La Palma in Spain 
   (runs 34-011 and 40-006) and the European Southern Observatory in Chile (run 075.C-0292).}}


  \author{P. P. Petrov\inst{1}
          \and G. F. Gahm\inst{2}
	        \and H. C. Stempels\inst{3}
	        \and F. M. Walter\inst{4}
	        \and S. A. Artemenko \inst{1}}

   \offprints{P. P. Petrov}

   \institute{Crimean Astrophysical Observatory, p/o Nauchny, Crimea, 98409 Ukraine\\
              email: \mbox{petrov@crao.crimea.ua}
              \and
              Stockholm Observatory, AlbaNova University Centre, Stockholm University,
	      SE-106\,91 Stockholm, Sweden
	      \and
	      Department of Physics and Astronomy, Uppsala University, Box 516, SE-751\,20 Uppsala, Sweden
	      \and
	      Department of Physics and Astronomy, Stony Brook University, Stony Brook, NY 11794-3800, USA	      }

   \date{}

   
 \abstract
  {Optical spectra of classical T Tauri stars (cTTS) are rich in emission lines
  of low-excitation species that are composed of narrow and broad components,
  which indicates the existence of two emitting regions with different
  kinematics, densities, and temperatures. The photospheric spectrum is often
  veiled by an excess continuous emission. This veiling is usually attributed
  to radiation from a heated region beneath the accretion shock. The broad
  emission lines of \ion{H}{i}, \ion{He}{ii}, \ion{Ca}{ii}, \ion{Fe}{ii}, and
  other species are thought to form in a larger volume of gas.} 
  {The aim of this research is to clarify the nature of the veiling, and
  whether the narrow chromospheric lines of \ion{Fe}{i} and other metals
  represent a standard chromosphere of a late-type star, or are induced by
  mass accretion.} 
  {We carried out high-resolution spectroscopy of selected cTTS with a special
  focus on DR Tauri and followed variations of chromospheric features, such as
  narrow \ion{Fe}{i} emission lines, and accretion signatures such as the
  veiling continuum and the \ion{He}{ii} line emission.}
  {We found that the amount of veiling in DR Tau varies from practically
  nothing to factors more than 10 times the stellar continuum intensity, and
  that the veiling is caused by both a non-photospheric continuum and
  chromospheric line emission filling in the photospheric absorption lines. The
  latter causes differential veiling because stronger lines are more
  veiled. We developed methods to separate the two sources of veiling. Several
  veiled T Tauri stars show a common effect: the radial velocities of
  photospheric and chromospheric lines vary in anti-phase. This is caused by an
  area with enhanced chromospheric emission, which is offset from the pole of
  rotation and is associated with the hot spot formed at the footprint of the
  magnetic funnel of mass accretion.}
 {The enhanced chromospheric emission in cTTS is linked not only to solar-like
  magnetic activity, but is powered to a greater extent by the accreting gas. 
  We suggest that the area of enhanced chromospheric emission is induced by
  mass accretion, which modifies the local structure of stellar atmosphere in
  an area that is more extended than the hot accretion spot. The narrow
  emission lines from this extended area are responsible for the extra
  component in the veiling through line-filling of photospheric absorption
  lines.}

 \keywords{stars: pre-main sequence -- stars: variables: T Tauri, Herbig Ae/Be -- Accretion, accretion disks -- stars: individual: DR Tau, S CrA, RW Aur, RU Lup, DI Cep}

 \maketitle
%

\section{Introduction}
\label{sec:intro}

The T Tauri stars are pre-main sequence stars of low mass. Spectroscopically
they are characterized by emission lines superimposed on a photospheric
spectrum of late type. In the classical T Tauri stars (cTTS) the emission line
spectrum is very rich and related to magnetospheric accretion from dusty
disks. The T Tauri stars rotate faster than most late-type main-sequence stars,
and the enhanced emission in the weak-line stars, those not exhibiting evidence
of accretion, is adequately explained in terms of enhanced solar-like magnetic
activity. The same dynamo effect should be present in the cTTS as well, but in
addition enhanced line emission is generated in the accreting gas, which flows
from the circumstellar disk to the star, where it eventually heats up and
releases energy in strong shocks at the stellar surface. Accordingly, the
emission lines are much broader in classical T Tauri stars than expected from
chromospheric emission alone. For reviews of properties and models of cTTS see
Petrov (\cite{petrov03}) and Bouvier et al. (\cite{bouvier07}). 

In cTTS the photospheric absorption lines are weakened by what is commonly
attributed to an overlying continuous excess emission. The energy distribution
of this {\it veiling} has been matched with black body radiation at a
temperature of $\sim 10^{4}$ K, or alternatively with Paschen continuous
emission (Basri \& Batalha \cite{basri90}; Hartigan et al. \cite{hartigan91},
\cite{hartigan95}; Valenti et al. \cite{valenti93}; Calvet \& Gullbring
\cite{calvet98}; Gullbring et al. \cite{gullbring98}). The veiling is
attributed to radiation from ``hot spots'', small areas (sometimes rings) of
heated gas at the footprints of the accretion flows. T Tauri stars are
photometrically variable, both from rotational modulation of surface features
and stochastic events (e.g., flares, changes in mass accretion rates). The
changes in line intensities and profiles in many cTTS are commonly attributed to
variations in the accretion flow. The veiling levels can vary dramatically, and
in the most active cTTS veiling may change from a tiny fraction of the intensity
of the photospheric continuum to factors of several times this level. The
degree of veiling is commonly used to estimate accretion rates in CTTS.

Intuitively, one would expect that when the veiling changes, the stellar
brightness would change accordingly. However, there had been several hints that
this is not always the case. Gahm et al.(\cite{gahm08}) demonstrated that in
four cTTS with very pronounced emission line spectra and strong irregular light
variations, the veiling variations clearly did {\it not} follow this pattern.
Substantial changes in the veiling were accompanied by only minor changes in $V$
and $B$, which in combination with an analysis of variations in equivalent
widths of emission lines led the authors to conclude that veiling is not solely
related to an excess continuum emission. Furthermore, it was demonstrated that
the photospheric absorption lines are weakened due to filling-in by narrow,
emission line components. While the above shows that the veiling effect is
caused by at least two different processes, it is still unclear which mechanism
causes the appearance of these narrow emission features.

In order to investigate the origin of the narrow-lined chromospheric emission
features, we performed a detailed investigation of the behavior of veiling in
one of the most variable cTTS, namely DR Tauri. Our observations show this star
to vary from a state of practically no veiling to levels of more than 10 times
the continuum flux. At high veiling levels narrow emission features are clearly
present in this object. In addition we find periodic variations in radial
velocity of both absorption lines as well as of the narrow emission components,
which, as we will show in this paper, provides indications on their origin. 

DR Tau is an unusually ``active'' cTTS, in the sense that the emission line
spectrum is exceedingly rich and variable, and that the photometric variations
are large and primarily related to physical changes in the plasma surrounding
the star. There are many other well-known cTTS with similar properties, for
which similar conclusions can be drawn regarding the nature of the veiling and
the radial velocity fluctuations. We will illustrate this in the present paper
by complementing our analysis with stars that show similar spectral line
variations, namely RW Aur A, DI Cep, RU Lup, and S CrA SE.


DR Tau was for a long period a relatively faint star until in 1976 it rather
suddenly brightened by several magnitudes in the blue (Chavarria-K.
\cite{chavarria79}), and the star is sometimes classified as an EXor star (see
e.g. Lorenzetti et al. \cite{lorenzetti09}). However, since that time it has
stayed at this bright level but with dramatic and fairly irregular light
variations (Artemenko et al. \cite{artemenko10}). Tentative photometric periods
proposed by Richter et al. (\cite{richter92}), Bouvier et al.
(\cite{bouvier93}), and Bouvier et al. (\cite{bouvier95}) differ by several
days, and are different from a more recent determination by Percy et al.
(\cite{percy10}), who assign a period of 5.0 days. This is also in the same
range of possible periods found from near-infrared (NIR) photometry by Kenyon et
al. (\cite{kenyon94}), who relate the periodicity of DR Tau to a hot spot on a
rotating star with a magnetic axis inclined to the stellar rotation axis.

Pronounced spectral variability in DR Tau was recognized early (Bertout et al.
\cite {bertout77}; Krautter \& Bastian \cite{krautter80}; Mundt \cite{mundt84}; 
Aiad et al. \cite{aiad84}; Appenzeller et al. \cite{appenzeller88}). The strong
emission lines on occasion show inverse P Cygni structures, which appear and
disappear on time-scales as short as a few hours (Smith et al.~\cite{smith97},
\cite{smith99}).  Basri \& Batalha (\cite{basri90}) demonstrated that the
veiling is large and variable, and subsequent  investigations of relations
between variations in emission lines and veiling led Guenther \& Hessman
({\cite{guenther93}) and Hessman \& Guenther (\cite{hessman97}) to conclude that
in principle the observations were consistent with the concept of magnetospheric
accretion, with gas flowing along dipole magnetic fields  to a hot region on the
visible surface. However, the data are inconsistent with constant accretion in a
simple  magnetic dipole, because irregular mass accretion and possible
flare-line events  distort the Balmer line profiles. In their long-term study of
spectral variablity in DR Tau, Alencar et al. (\cite{alencar01}) were not able
to recover any periodicity in the spectral changes. However, Hessman \& Guenther
(\cite{hessman97}) found evidence of a periodicity of 4.5--5 days in the
equivalent widths of strong lines, and  Johns \& Basri (\cite{johns95}) reported
a 5.1 day period in components in H$\alpha$. Because these spectral periods are
similar to some of the photometric periods presented above, one may suspect that
both phenomena are related to rotational modulation, and that the period of
rotation of DR Tau is in the range of 4.5--5 days.

The complex emission line profiles and their dramatic changes can be better
understood if the profiles are decomposed into different components arising from
different emitting volumes around the star. The accreting gas produces Balmer
line emission, sometimes with inverse P Cygni absorption components, and the
broad metallic lines reflect the bulk motion of this in-falling gas (see e.g.
Muzerolle et al. \cite{muzerolle98a}, \cite{muzerolle98b}). Normal P Cygni
absorption components are seen as well, indicating outflow in the form of
massive winds or jets. The P Cygni absorptions are particularly distinct in some
far-ultraviolet lines, like the \ion{Mg}{ii} doublet (Kravtsova \& Lamzin
\cite{kravtsova96}, Ardila et al. \cite{ardila02}). The \ion{He}{i} lines,
particularly the NIR line at 10830\,\AA, display absorption both from
the in-falling and out-flowing gas as analysed in a series of papers by
Beristain et al. (\cite{beristain01}), Edwards et al. (\cite{edwards03},
\cite{edwards06}), Kwan et al. (\cite{kwan07}) and Fischer et al.
(\cite{fischer08}). In all, DR Tau stands out as a cTTS with an unusually
dynamic and very complex interplay between accreting and ejected gas.

As in many cTTS, the He lines in DR Tau display a narrow emission component,
which normally is attributed to post-shock gas at the footprints of the
accretion flows  (Lamzin et al.\cite{lamzin96}, Beristain et al.
\cite{beristain98}). However, in DR Tau similar narrow components with a
full-width half-maximum (FWHM) of $\sim$ 20 km s$^{-1}$ were also found in lines
of \ion{Ca}{ii}, \ion{Fe}{i} and \ion{Fe}{ii} (Batalha et al. \cite{batalha96}, 
Beristain et al. \cite{beristain98}). In fact, Batalha et al. found a positive 
correlation between line strength and veiling and speculated that the narrow 
component is related to a heated stellar atmosphere close to the accretion zone
--  a ``hot chromosphere''.

Below we present our observations and reduction methods in
Section~\ref{sec:obsred}. The methods of analysis and results are given in
Section~\ref{sec:analysis} and are discussed in more detail
Section~\ref{sec:discussion}. Section~\ref{sec:conclusions} summarizes our
results.

\section{Observations and data reductions}
\label{sec:obsred}

Observations of DR Tau were carried out during two runs in 2007 and 2009/2010 at
the Nordic Optical Telescope (NOT) with the fiber-fed echelle spectrograph FIES
(Frandsen \& Lindberg \cite{frandsen00}). The data were processed with
FIEStool\footnote{
http://www.not.iac.es/instruments/fies/fiestool/FIEStool.html}. This software
package is fine-tuned for reducing data obtained with the FIES spectrograph and
performs the full set of echelle spectral reduction steps, including bias
subtraction, modelling of scattered light, extraction through two-dimensional
modelling of the spectral orders, flat-fielding and wavelength calibration.

The extracted spectra have a resolution of $\lambda/\Delta\lambda \approx
47\,000$ and cover a wavelength range of $3650$--$7250$ {\AA}. In
Table~\ref{obs_log} we summarize observation dates, the number of 20-min
exposures obtained in each night and signal-to-noise ratio per pixel in the
continuum of nightly averaged spectra. The last two columns give the mean
veiling factors at 6110--6160\,\AA\, and their standard
deviations (see Sect. 3.3.1). In addition, we obtained spectra of the K7V flare
star HD 28343 (BD +21$\degr$ 652) for comparison purposes.

\begin{table}
\caption{DR Tau observations. See the main text for a more detailed description of the columns.}
\label{obs_log}
\begin{center}
\begin{tabular}{c|c|c|c||c|c}
\hline
\hline
 Year & HJD 245... & N exp & S/N ratio & VF$_{6110}$ & $\sigma_{VF}$ \\
\hline
2007 & 4101.4056 & 2 &  65 & 4.4 & 0.5\\   
     & 4103.5857 & 2 &  45 & 1.5 & 0.3\\  
     & 4104.4278 & 2 &  55 & 1.2 & 0.3\\   
     & 4105.6235 & 2 &  80 & 7.5 & 1.0\\   
2009 & 5192.5623 & 4 & 110 & 4.8 & 0.5\\  
     & 5193.4952 & 3 &  80 & 7.0 & 1.0\\   
     & 5194.4984 & 3 &  60 & 5.5 & 0.5\\  
     & 5195.4934 & 4 &  90 & 3.8 & 0.5\\  
     & 5196.4748 & 4 &  70 & 4.6 & 1.0\\  
2010 & 5197.5002 & 4 & 150 & 7.0 & 0.7\\  
     & 5198.5539 & 4 &  90 & 4.6 & 0.5\\  
     & 5199.5762 & 2 &  35 & 5.0 & 1.5\\ 
     & 5201.4676 & 2 &  60 & 3.6 & 0.5\\  
\hline
\end{tabular}
\end{center}
\end{table}

This study focuses on DR Tau, but includes also supporting data from the two
components of S CrA (Gahm et al. \cite{gahm08}), RU~Lup (Stempels et al.
\cite{stempels07}), RW~Aur~A (Petrov et al. \cite{petrov01}), and DI~Cep
(Gameiro et al. \cite{gameiro06}).

\section{Data analysis and results}
\label{sec:analysis}

We start with a brief description of the spectra of DR Tau and the
determination of the basic stellar parameters: temperature, gravity, radius, and
rotation, and continue with an analysis of the veiling as a tracer of
accretion, and radial velocities of narrow emission lines as indicators of
chromospheric surface inhomogeneities. 

\subsection{General spectral properties}

We present in Fig.~\ref{dr_emiss} two typical segments of the DR Tau spectrum.
Here, we averaged all 13 spectra to increase the signal-to-noise ratio. For
comparison we also show a spectrum of the template K7 star artificially  veiled
by a factor of four, so that the strength of the photospheric lines become
similar to those of DR Tau.

\begin{figure}
\centerline{\resizebox{9cm}{!}{\includegraphics{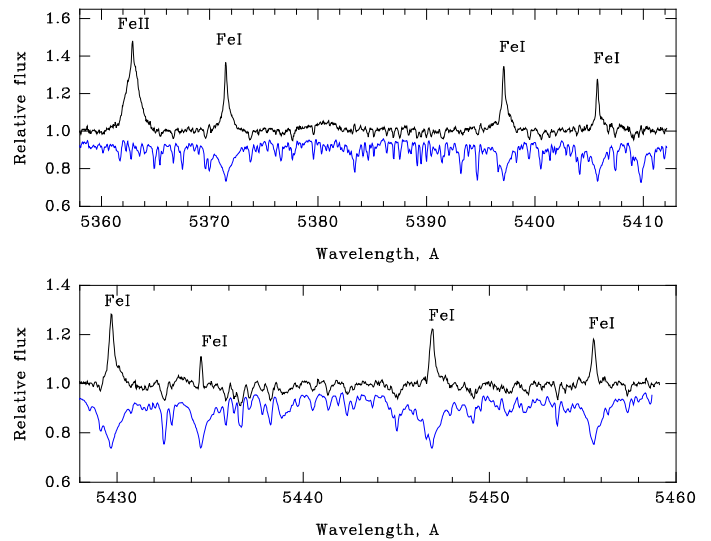}}}
\caption{Segments of the averaged spectrum of DR Tau (solid line) as compared to
the K7 V  template HD 28343 (thin line) veiled artificially by a factor of four.
In this and other figures the  wavelength scale is in the stellar rest frame.}
\label{dr_emiss}
\end{figure}

The main features in the spectrum of DR Tau are as described by Beristain et al.
(\cite{beristain98}). The strongest emission is in the hydrogen Balmer lines and
\ion{Ca}{ii} H and K. Weaker emission is seen in neutral and singly ionised
permitted lines of \ion{Fe}{i}, \ion{Fe}{ii}, \ion{He}{i}, \ion{He}{ii},
\ion{Ti}{ii}, \ion{Na}{i} and \ion{Ca}{i}, and in forbidden [\ion{O}{i}],
[\ion{S}{ii}], and [\ion{N}{ii}].

The weakest emission lines are narrow (FWHM $\sim$ 20 km\,s$^{-1}$), while
stronger emission lines of metals exhibit a two-component structure: a similar
narrow component as well as a broad component with FWHM $\sim$ 100 km\,s$^{-1}$.
The broad component gradually grows stronger as the emission lines themselves
become stronger, as is evident from the panels shown in Fig.~\ref{dr_emiss}. The
stronger \ion{He}{i} and \ion{Fe}{ii} lines (for example \ion{Fe}{ii} $\lambda$
4924 and $\lambda$ 5018\,\AA) show additional red-shifted absorption extending
to +350 km\,s$^{-1}$. These signatures of infall vary with time and are also
present in the higher Balmer lines as shown in Fig.~\ref{dr_accr}. The profiles
of H$\alpha$ and other Balmer lines are split by deep broad absorption, centred
at $-100$ km\,s$^{-1}$, which indicates an outflow. Typical Balmer line profiles
can be found in Alencar et al. (\cite{alencar01}), where the Balmer emission was
analysed in greater depth.

\begin{figure}
\centerline{\resizebox{9cm}{!}{\includegraphics{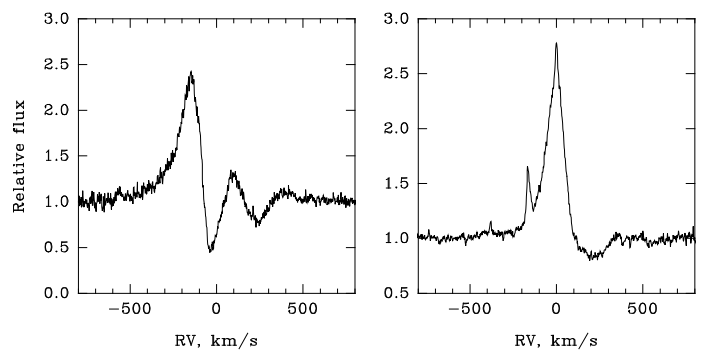}}}
\caption{Emission line profiles with red-shifted absorption components in the spectrum of DR Tau on HJD 2455192. Left: H$\delta$, right: \ion{Fe}{ii} 5018.43\,\AA\, (the narrow emission peak on the blue wing is \ion{He}{i} 5015.67\,\AA).}
\label{dr_accr}
\end{figure}

Photospheric lines are detectable longward of $\sim$ 4500\,\AA\, in the averaged
spectrum, while during some nights with higher veiling and/or lower
signal-to-noise ratio many lines are not detectable. Nevertheless, in some
spectral intervals the photospheric lines are always visible, which makes it
possible to follow night-to-night variations of the veiling. In moments of high
veiling some photospheric lines (e.g. \ion{Fe}{i}, multiplet 15)  turn from
absorption into pure emission.

\subsection{Stellar parameters}

Spectral types assigned previously for DR Tau fall in a fairly narrow range,
from K5 to M0. The variable and often high veiling of the photospheric
absorption lines makes it difficult to derive precise stellar parameters.
However, we did obtain one spectrum of DR Tau in a relatively calm state (HJD =
2454104), which we will refer to as the low-veiling spectrum.

This low-veiling spectrum was compared to a grid of synthetic spectra calculated
for $T_{\rm eff}$ = 3500, 3750, 4000, 4250, and 4500 K with log $g$ values of
3.0, 3.5, and 4.0. This grid was calculated using the code by Berdyugina
(\cite{berdygina91}) and Kurucz' models. Atomic line data were retrieved from
the VALD database (Kupka et al. \cite{kupka00}). The best fit was found for
$T_{\rm eff}$ = 4000 -- 4250 K and log $g$ = 3.5 -- 4.0. In order to check the
spectral type with molecular features, we used the spectrum of HD 28343
(spectral type K7V) along with M0V and K5V spectra collected earlier with FIES.
The spectrum of DR Tau is best matched with the K7V spectrum, as is illustrated
in Fig.~\ref{tio_7054}, where the template spectrum is artificially veiled by a
factor of 1.0. The TiO bandhead at 7054\,\AA\, is also consistent with
a spectral type of K7V for DR~Tau, and we adopt $T_{\rm eff}$ = 4100 $\pm$ 200 K
and log $g$ = 3.7 $\pm$ 0.5 for the photosphere of DR Tau.

In Fig.~\ref{dr_synt} we compare the low-veiling spectrum of DR Tau with a
synthetic spectrum  in the region of  \ion{Ca}{i} $\lambda$ 6124\,\AA. The
synthetic spectrum was calculated for  $T_{\rm eff}$ = 4100 K, log $g$ = 3.7, 
$v_{\rm mic}$ = 1.5\,km\,s$^{-1}$,  $v_{\rm mac}$ = 1.5\,km$\,s^{-1}$ and
$v\,\sin i$ = 5 km\,s$^{-1}$,  convolved with the instrumental profile of FIES
and veiled by factor of 1.3.  The profiles of the \ion{V}{i}, \ion{Ti}{i} and
\ion{Ba}{ii} lines fit fairly well.  A $\log g$ in the range 3.5--4.0 gives a
reasonably good fit to the broad wings of  the \ion{Ca}{i} line. The lines of
\ion{Fe}{i} are partly filled in with chromospheric  emission (see below). }

The projected rotational velocity $v\,\sin i$ = 5 $\pm$ 1 km\,s$^{-1}$ was found
from  the widths of many metal lines as fitted with the synthetic spectrum  in
the region 5580--6150\,\AA. Note that $v\,\sin i$ is somewhat
less than  the FWHM of the instrumental profile ($\approx$6.5 km\,s$^{-1}$).
Nevertheless, the high signal-to-noise ratio in our spectra enables us to detect
this low value of the rotational broadening of photospheric lines.

Assuming that at minimum brightness the photosphere is less affected by
accretion, we can estimate the stellar luminosity and radius.  From an extended
set of photometric data from 1986--2003 (Grankin et al. \cite{grankin07}), 
$V_{\rm min} = 13 \fm 0 \pm 0 \fm 30$ and the corresponding colour $V-R = 1 \fm
40 \pm 0 \fm 08$, which gives $A_V = 0 \fm 93 \pm 0 \fm 30$. From the
corresponding minimum $B-V$ colours we find a faintest $B$ magnitude that is
similar to what DR Tau had in the first half of the 20th century (Chavarria-K.
\cite{chavarria79}).  Hence, there is no indication that the stellar parameters
have changed following the general increase in brightness around 1976. The
circumstellar component of the extinction is probably small because the star is
seen close to pole-on (see also the discussion in Sect.
\ref{sec:threemethods}). With a distance to DR Tau of $d$ = 140 pc (e.g.
Bertout \& Genova \cite{bertout06}) we obtain the following stellar parameters:
$L = 0.52 \pm 0.15\,{\rm L}_{\sun}$ and $R=1.46 \pm 0.20\,{\rm R}_{\sun}$. 
Earlier published estimates of the radius range from $1.2\,{\rm R}_{\sun}$ 
(Bertout \& Basri \cite{bertout88}) to $2.7\,{\rm R}_{\sun}$ (Johns-Krull \&
Gafford \cite{johns02}).


Regarding other programme stars, the primary and secondary of S CrA have earlier
been classified as G5 Ve and K5 Ve, respectively (Carmona et al.
\cite{carmona07}). However, our long-slit UVES/VLT spectra yield very similar
parameters for the two components,  namely $T_{\rm eff}$ = 4250 $\pm$ 150 K,
$\log g$ = 4.0 $\pm$ 0.2 and $v\,\sin i$ = 12 $\pm$ 1 km\,s$^{-1}$. 
Furthermore, RW Aur A has a spectral type of K1--K4, $v\,\sin i \approx 20$
km\,s$^{-1}$ (Petrov et al.~\cite{petrov01});  RU Lup has $T_{\rm eff}$ = 3950
$\pm$ 200 K, $\log g$ = 3.9 $\pm$ 0.3, and $v\,\sin i$ = 9.0 $\pm$ 0.9
km\,s$^{-1}$ (Stempels \& Piskunov~\cite{stempels02}), and DI Cep has spectral
type G8 and $v\,\sin i$ = 20.0 $\pm$ 1.5 km\,s$^{-1}$ (Gameiro et
al.~\cite{gameiro06}).

\begin{figure}
\centerline{\resizebox{8cm}{!}{\includegraphics{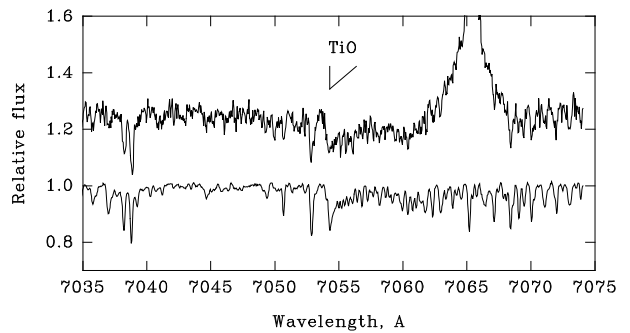}}}
\caption{Molecular TiO band at 7054\,\AA. Upper: the low-veiling spectrum of DR~Tau. Lower: the spectrum of HD~28343, (K7~V), artificially veiled by a factor of 1.0.}
\label{tio_7054}
\end{figure}

\begin{figure}
\centerline{\resizebox{9cm}{!}{\includegraphics[width=.9\textwidth]{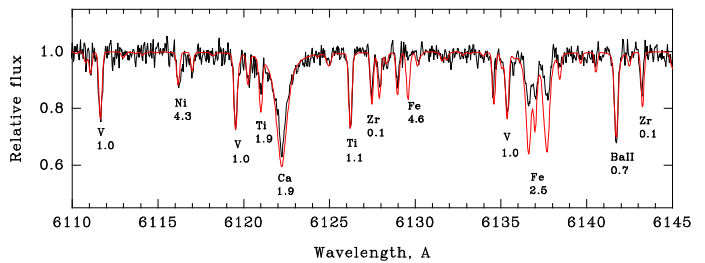}}}
\caption{Solid line: the low-veiling spectrum of DR~Tau. Thin line: synthetic
spectrum for $T_{\rm eff}$ = 4100~K, log $g$ = 3.7 and $v\,\sin i$ = 5\,
km\,s$^{-1}$, convolved with the instrumental profile and veiled by a factor of
1.3. Elements and lower level excitation potentials (eV) are indicated.}
\label{dr_synt}
\end{figure}

\subsection{Two sources of veiling}

As discussed early by e.g. Bertout (\cite{bertout84}), the photospheric spectrum
of a cTTS can be veiled by 1) an additional non-photospheric excess continuum
and 2) line emission filling in photospheric absorption lines. 

\subsubsection{Traditional veiling measurements}

Traditionally, the veiling factor (VF) is determined from the equivalent width
(EW) of a photospheric line measured in the spectrum of DR Tau and in a
template (e.g. synthetic) spectrum: 

     $${\rm VF} = {\rm EW}_{\rm synt}/{\rm EW}_{\rm DR} - 1.$$

\begin{figure}
\centerline{\resizebox{8cm}{!}{\includegraphics{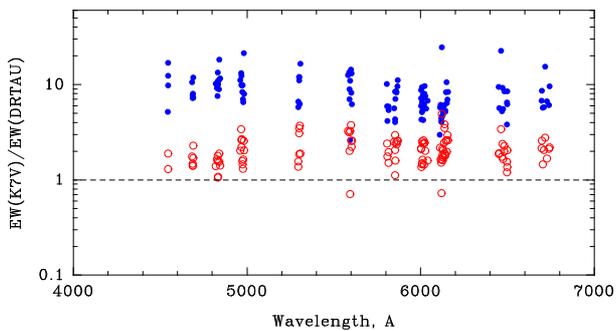}}}
\caption{Ratio of EWs in the K7V template and DR Tau. Open circles: low-veiling
spectrum, filled circles: average spectrum. The horizontal dashed line marks the
case of zero veiling.}  
\label{lam_vei3}
\end{figure}

From measurements of numerous photospheric absorptions in the average spectrum
of DR Tau no significant correlation was found either in veiling versus
wavelength, or in veiling versus excitation potential of different lines, within
the wavelength range 4500--6800\,\AA\, and excitation
potential range 0--5 eV. Although it is commonly accepted that veiling increases
towards the blue, the wavelength dependence is not always well defined because
of the large scatter in the veiling factors derived from different spectral
lines (e.g. Basri \& Batalha, \cite{basri90}). In Fig.~\ref{lam_vei3} we show
the ratio of EWs as measured in the spectra of the K7V template and DR Tau as a
function of wavelength. Each symbol represents one spectral line. Note that the
large scatter of points is caused not only by errors of measurements of EWs, but
also by the dependence of veiling on line strength (see below).

Not all absorption lines that are visible in the averaged spectrum can be
detected in {\it individual} spectra. We therefore carefully selected spectral
regions where photospheric lines are clearly visible in {\it all} spectra of DR
Tau, even in those with high veiling, and which never turn into emission:
5580--5605\,\AA\, (six lines of \ion{Ca}{i}),
6000--6040\,\AA\, (four lines of \ion{Mn}{i} and
\ion{Fe}{i}), 6110--6160\,\AA\, (five lines of \ion{V}{i},
\ion{Ti}{i}, \ion{Ba}{ii}, \ion{Na}{i}) and 6460--6480\,\AA\,
(two lines of \ion{Ca}{i}). In these regions we measured the corresponding
values of the VF in each spectrum of DR Tau using the K7V template.
Fig.~\ref{tim_veil} shows the night-to-night variations of the veiling. The VF
at 6110--6160\AA\, and its standard deviation are also given
in the last two columns in Table~\ref{obs_log}.

\subsubsection{Distinguising the excess continuum from chromospheric line
emission}
\label{sec:threemethods}

Unfortunately, the veiling measurements we perform above cannot discriminate
between the effect of a veiling {\it continuum} and the effect of chromospheric
{\it line emission} filling-in the photospheric lines. We therefore used
the following three methods to separate and quantify these two effects.

\begin{figure}
\centerline{\resizebox{7cm}{!}{\includegraphics{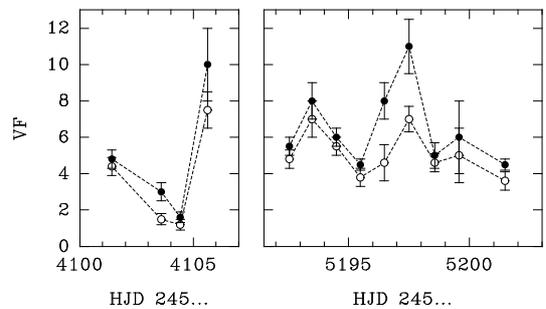}}}
\caption{Night-to-night variations of the veiling factor in DR Tau. Filled
circles: spectral region 6000--6040\,\AA\, (lines of \ion{Mn}{i} and
\ion{Fe}{i}). Open circles: spectral region 6110--6160\,\AA\, (lines
of \ion{V}{i}, \ion{Ti}{i}, \ion{Ba}{ii}, \ion{Na}{i})}
\label{tim_veil}
\end{figure}

The first method is based on measurements of the forbidden lines and is similar
to the method used by Gahm et al. (\cite{gahm08}). Forbidden emission lines of
[\ion{O}{i}] at 5777.3, 6300.2 and 6363.8\,\AA\,
and [\ion{S}{ii}] at 6717.0 and 6731.3\,\AA\, are usually
present in spectra of cTTS. These forbidden lines are formed in a large volume
extending far away from the star, as compared to permitted emission
lines, and can therefore be expected to be intrinsically stable on a time scale
of days. Hence, the observed night-to-night variations of EWs of the forbidden 
lines are mainly caused by {\it continuum} variations.

\begin{figure}
\centerline{\resizebox{6cm}{!}{\includegraphics{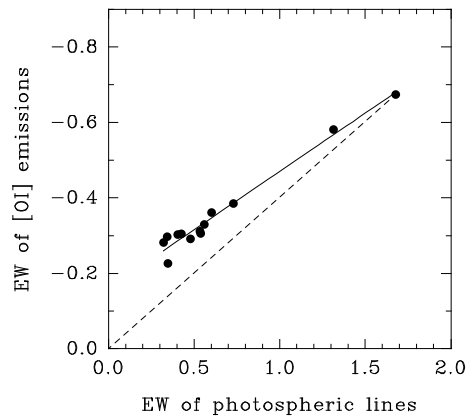}}}
\caption{Correlated variations of EWs of forbidden emission lines (sum of
[\ion{O}{i}] $\lambda$ 5577,  $\lambda$ 6300 and $\lambda$ 6363\,\AA) and
photospheric lines (sum of 17 lines within 5580--6500 \AA). Dashed line shows
relation expected in case of only continuum variations.}
\label{o1_veil}
\end{figure}

Fig.~\ref{o1_veil} shows that there is a strong correlation between EWs of
photospheric absorption lines and forbidden lines of [\ion{O}{i}]. The existence
of this correlation indicates that both the photospheric and the [\ion{O}{i}]
lines are affected by the continuous veiling component. However, the range of
variability for the photospheric lines is twice as large as for the [\ion{O}{i}]
lines: the EWs of [\ion{O}{i}] lines change by a factor of 2.7, while the EWs of
photospheric lines change by a factor of of 5.3. This means that another source
of veiling contributes as much as the continuum veiling component.

The second method is based on the effect of differential veiling. Here, we
measured from the average spectrum of DR Tau EWs of about 90 photospheric lines 
of different elements in the spectral region 5580--6720\,\AA.
We then measured the same lines in the template spectrum. For each spectral line
we calculated the ratio of EWs in DR Tau and in the template spectrum. Identical
measurements were performed on the low-veiling spectrum of DR Tau.

\begin{figure}
\centerline{\resizebox{8cm}{!}{\includegraphics{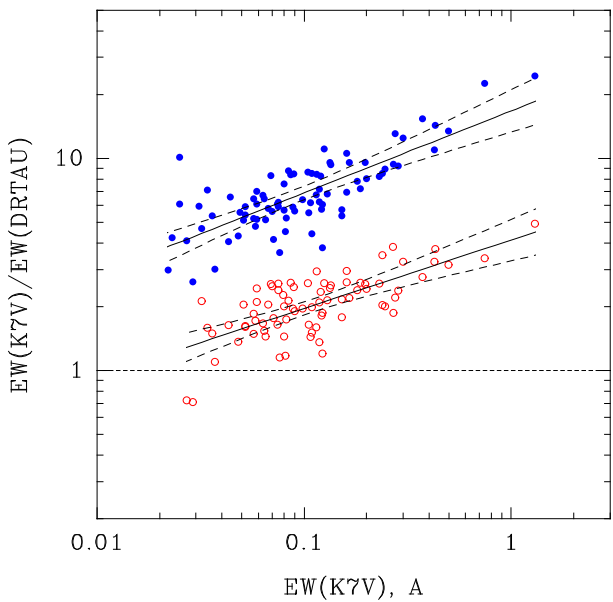}}}
\caption{Ratio of EWs of photospheric lines of DR Tau and the observed K7V
template HD 28343 as a function of line strength. Filled circles -- averaged
spectrum of DR Tau. Open circles -- low-veiling spectrum of DR Tau. Regression
lines are shown for both sequences. The 99\% confidence intervals are outlined
by dashed curves. The scatter of points around the regression lines is caused by
errors of measurements of EWs. The horizontal dashed line corresponds to zero
veiling.}

\label{vei_dep}
\end{figure}

If the veiling were {\it only} caused by a featureless continuous veiling
component, all lines would be reduced by the same factor. However, the ratio of
EWs depends on line strength: {\it stronger lines are veiled more} (see
Fig.~\ref{vei_dep}). The effect as such has been recognized before (e.g. Basri
\& Batalha \cite{basri90} and references therein) but is here demonstrated in a
graphical form. We obtained the same result when we used a synthetic template
spectrum instead of the K7V data, and the effect of the differential veiling
persists when temperature and gravity of the template vary within probable
errors in $T_{\rm eff}$ ($\pm$ 200 K) and log $g$ ($\pm$ 0.5). Note that the
scatter of points in Fig.~\ref{vei_dep} is solely owing to uncertainties in the
EW measurements from the spectra of DR Tau and the template. From the weakest
lines (EW $\leq$ 0.05\,\AA) in Fig.~\ref{vei_dep} we estimate VF $\sim$ 0--0.5
for the low-veiling spectrum, and VF $\sim$ 3--4 for the averaged spectrum.
These numbers represent upper levels of the veiling {\it continuum} in the two
spectra. The increasing excess of veiling towards stronger lines (EW $>$
0.1\,\AA) should be attributed to chromospheric {\it line emission}. The VF
values given in Table~\ref{obs_log} and shown in Fig.~\ref{tim_veil} are derived
from relatively strong photospheric lines (EW $\approx$ 0.1\,\AA\, in the K7V
template), because weaker lines are not always measurable in the individual
spectra of DR Tau.

The third method uses the colourimetric characteristic of a star with a hot
spot  (Petrov \& Kozack \cite{petrov07}). According to standard magnetospheric
accretion models the continuum veiling originates from the emission of a hot
spot located below the accretion stream. The expected effective temperature of
the hot spot is 6000--8000 K with a filling factor within 0.1--1 \% of the
stellar surface area (Calvet \& Gullbring \cite{calvet98}). It is believed that
stochastic events of accretion cause the irregular night-to-night variability of
the brightness and colours of cTTS.

The expected colours of a K7 star with a hot photospheric spot can be calculated
as a function of spot size and temperature. These two parameters define the
veiling factor VF, and hence the calculated colours. This means that, for
instance, changes in $B-V$ and $V-R$ are directly related to the degree of
veiling. In order to analyse the colour changes in DR Tau we made use of the
photometric catalogue by Grankin et al. (\cite {grankin07}), which includes
about 20 years of observations of TTS. For DR Tau there are over 600
observations in 18 seasons from 1986 to 2003.

\begin{figure}
\centerline{\resizebox{9cm}{!}{\includegraphics{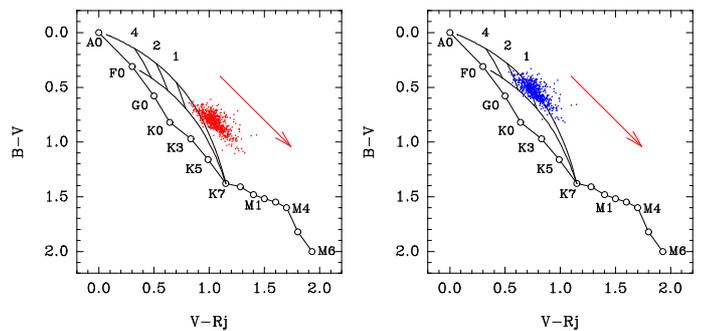}}}
\caption{Two-colour diagrams with a grid of ''star + spot'' models, 
shown with the bow-shaped curves.
The numbers along the curves represent the amount of veiling in each model.
In the left panel, the cluster of dots are the observed colours of DR Tau. 
In the right panel, a correction for interstellar reddening $A_V = 1\fm 0$ has been applied. 
The slope of the interstellar reddening is indicated with an arrow.}
\label{dr_hotk7}
\end{figure}

The left panel of Fig.~\ref{dr_hotk7} shows colour-colour diagrams with the main
sequence from A0 to M6, as well as the observed colours of DR Tau. The two
solid bow-shaped curves are sequences of the ``star+spot'' colours,
corresponding to two different {\it colour temperatures} of the hot spot,
approximately those of F0 and A0 stars. Along these curves, the spot radius
varies, thus changing the filling factor and, correspondingly, the veiling
factor VF. The loci of VF = 1, 2, and 4 are marked in the figure. The observed
colours deviate from the model grids, but corrections for interstellar reddening
must be made.

The right panel of Fig.~\ref{dr_hotk7} shows the same diagram with the colours
of DR Tau corrected for an interstellar reddening $A_V = 1\fm 0$. This
correction brings the cluster of points to the sequence of the ''star + spot''
colours. The cluster of points is centred at the position of VF $\approx$ 1.0,
with a spread from VF = 0.5 to VF = 2. Some scatter towards bluer B-V colours
(about $0\fm 2$) may be caused by contribution of emission lines to the B pass
band. 

Apparently, the hot region is permanently present on the visible surface of the
star, but does not contribute to the veiling by more than VF $\approx$ 2.0. 
This corresponds to a brightness variability in the V band within $\approx$
1$^m$.2. Because the observed veiling of DR Tau reaches much larger values,
typically 3 to 10, the rest of the veiling effect must be caused by the line
emission filling-in the photospheric lines. Generally, if the effect of the line
emission reduces the photospheric line depth by a factor of VF$_{\rm line}$ and,
in addition, the non-photospheric continuum reduces the line depth by a factor
of VF$_{\rm cont}$, the resulting veiling VF is defined by the relation

   $$({\rm VF} + 1) = ({\rm VF}_{\rm line} + 1)\cdot({\rm VF}_{\rm cont} + 1).$$

This is a steep relation. For instance, with VF$_{\rm line}$ = VF$_{\rm cont}$ =
2.0, the resulting veiling is VF = 8. This explains the extremely high veiling
(VF $ > $ 10) observed sometimes in spectra of cTTS. It also means that the
accretion rate derived on the assumption that the observed veiling is caused by
an additional {\it continuum} may be overestimated by a considerable factor, as
was also concluded earlier by Gahm et al. (\cite{gahm08}) for other cTTS.

Note that the temperature used in this method is a {\it colour} temperature in
the spectral range covering the BVR pass bands, and may differ from the
effective temperature. The continuum radiation from a hot spot on a cTTS is
composed of an optically thick continuum from the heated photosphere and an
optically thin Paschen continuum from the pre- and post-shocked gas (Gullbring
et al. \cite{gullbring00}).

We conclude that all three methods described above show that the veiling {\it
continuum} in DR Tau is relatively small compared to the values  derived
directly from measurements  of EWs of the photospheric lines, indicating that
the excess of the observed veiling is caused by the chromospheric line
emission.

\subsection{Anti-phase radial velocity variations} 

Stellar radial velocity (RV), as measured from positions of photospheric lines,
can vary on time scales of days either because of orbital motions in a close
binary system or because of surface brightness inhomogeneities (cool/hot
spots). In the latter case variations in RV must be within the observed
$v\,\sin i$ value. 

Stellar RVs of DR Tau were measured by cross-correlation with the spectrum of
the template HD 28343 in the region 5560--6670 \,\AA, with the
emission lines masked out. The radial velocity of the template was determined
by cross-correlation with the synthetic spectrum ($T_{\rm eff}$ = 4000 K, log g
= 4.5) resulting in RV = $-35.20$ $\pm$ 0.04 km\,s$^{-1}$. The results of the RV
measurements are given in Table~\ref{rad_vel}.

\begin{table*}
\caption{Radial velocities of photospheric and chromospheric lines in DR Tau}
\label{rad_vel}
\begin{center}
\begin{tabular}{c|c|c}
\hline
\hline
   HJD-245... &   RV$_{\rm phot}$  & RV$_{\rm em}$\\
              &   (km\,s$^{-1}$)   & (km\,s$^{-1}$)\\ 
\hline   
   4101.4056  &   21.53 $\pm$0.35 & 22.12 $\pm$0.34\\
   4103.5857  &   23.08 $\pm$0.33 &  (...)  \\
   4104.4278  &   22.97 $\pm$0.15 &  (...)  \\
   4105.6235  &   23.14 $\pm$0.37 & 21.62 $\pm$0.23\\
   5192.5623  &   23.36 $\pm$0.18 & 22.19 $\pm$0.17\\
   5193.4952  &   23.08 $\pm$0.36 & 23.39 $\pm$0.31\\
   5194.4984  &   22.78 $\pm$0.29 & 23.24 $\pm$0.24\\
   5195.4934  &   22.98 $\pm$0.19 & 22.24 $\pm$0.18\\
   5196.4939  &   23.78 $\pm$0.25 & 21.70 $\pm$0.22\\
   5197.5002  &   23.31 $\pm$0.23 & 22.28 $\pm$0.28\\
   5198.5539  &   22.69 $\pm$0.15 & 24.00 $\pm$0.19\\
   5199.5762  &   22.96 $\pm$0.47 & 22.36 $\pm$0.19\\
   5201.4676  &   23.30 $\pm$0.21 & 22.06 $\pm$0.32\\
\hline
\end{tabular}
\end{center}
\end{table*}

\begin{table*}
\caption{Mean RV and range of RV variations}
\label{mean_rv}
\begin{center}
\begin{tabular}{r|c|c}
\hline
\hline
    Line     &      Mean RV       &    Semi-amplitude\\ 
             &    (km\,s$^{-1}$)  &    (km\,s$^{-1}$)\\
\hline             
    Photospheric lines & 23.0 $\pm$0.2 &  1.1 $\pm$0.2\\
\hline    
    Narrow component:  \ion{Fe}{i}  & 22.5 $\pm$ 0.2 &  1.1 $\pm$ 0.2\\
                      \ion{Fe}{ii}  & 23.4 $\pm$ 0.3 &  1.5 $\pm$ 0.2\\
                     \ion{He}{i}    & 26.9 $\pm$ 0.9 &  1.1 $\pm$ 0.3\\
                     \ion{He}{ii}   & 36.1 $\pm$ 0.5 &  2.5 $\pm$ 0.7\\
                     
\hline        
    Broad component: \ion{Fe}{i}, \ion{Fe}{ii}  & 24.3 $\pm$ 1.0  & 5.0 $\pm$1.5 \\  
                     \ion{He}{i}  & 22.9 $\pm$ 1.0 & 8 $\pm$2 \\     
\hline
\end{tabular}
\end{center}
\end{table*}

\begin{figure}
\centerline{\resizebox{8cm}{!}{\includegraphics{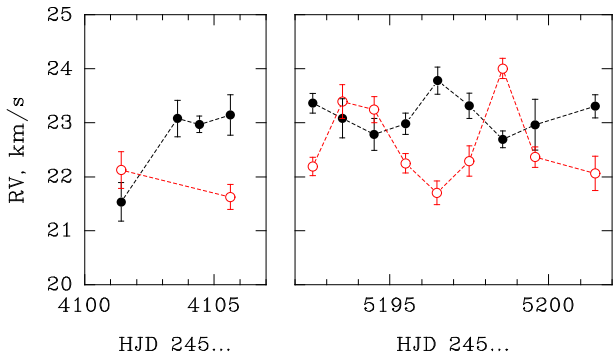}}}
\caption{Anti-phase RV-variations of photospheric (filled circles) and chromospheric (open circles) lines in DR Tau.}
\label{rv_abs}
\end{figure}

We then selected a group of unblended emission lines of the \ion{Fe}{i}
multiplet 15,  within 5371--5456\,\AA, as representative of
the chromospheric emission (most lines are present in Fig.~\ref{dr_emiss}). The
profile of each line was decomposed into a broad and narrow component, and the
position of the narrow component was determined by a Gaussian fit. The average
RVs of the \ion{Fe}{i} emission lines are also given in Table~\ref{rad_vel}. The
two blanks in the column of RV$_{\rm em}$ correspond to dates when the star was
at a state of low veiling and the \ion{Fe}{i} emission was too weak to be used
for RV measurements. The RVs show night-to-night variations with a
semi-amplitude of about 1 km\,s$^{-1}$ for both the photospheric and the
emission lines, which is well within the $v\,\sin i$ value. {\it The main result
is that the RV variations of the photospheric lines and of the narrow component
of the \ion{Fe}{i} emissions are in anti-phase} (see Fig.~\ref{rv_abs}). Similar
variations were first reported for RW Aur by Petrov et al. (\cite{petrov01}).

\begin{figure}
\centerline{\resizebox{9cm}{!}{\includegraphics{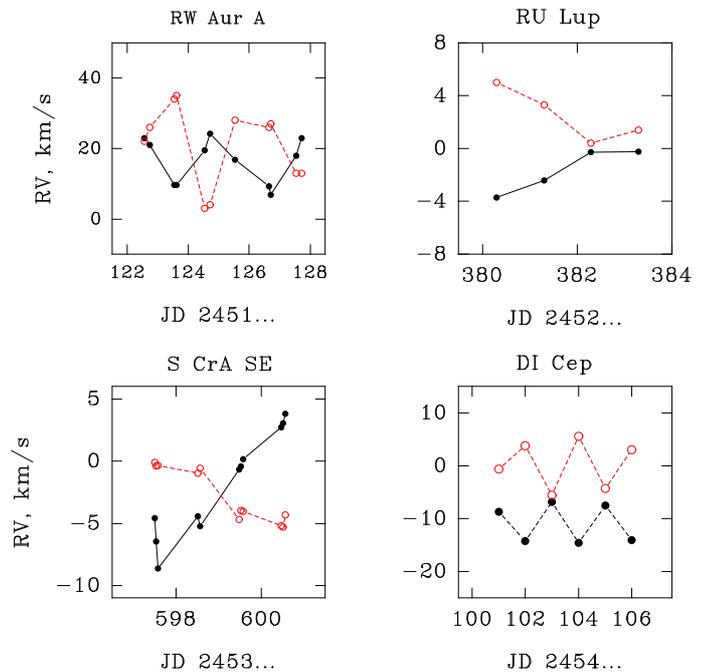}}}
\caption{Anti-phase RV-variations of photospheric (filled circles) and
chromospheric (open circles) lines in four cTTS. Narrow emission lines of
\ion{Fe}{i} were measured in RW Aur and S CrA SE, and of \ion{He}{i} in RU Lup
and DI Cep. Note the positive shift in velocity of the \ion{He}{i} emission.}
\label{anti-ph0}
\end{figure}

The broad components show periodic shifts in central velocity with time.
However, the measurement accuracy of the position of a broad component using a
Gaussian fit is much lower than for the narrow components. The RV {\it
amplitudes} of broad components were measured by means of cross-correlation of
each spectrum with the average spectrum of DR Tau with the narrow components
masked out. The amplitudes of the broad components turned out to be larger than
those of the narrow components, which is consistent with an origin of the broad
components well above the stellar surface.

Table~\ref{mean_rv} summarises measurements of RVs in different groups of lines,
which are arranged in order of increasing RV amplitude. The mean stellar
velocity is +23 km\,s$^{-1}$. Note that the mean RV is more positive for the
''high-temperature'' lines, up to +36 km\,s$^{-1}$ for the narrow \ion{He}{ii}
component. 

Our search for radial velocity changes was then extended to the other cTTS
included in the present study. As shown in Fig.~\ref{anti-ph0}, all programme
stars show the same type of anti-phase variations as observed for DR Tau.
Furthermore, narrow emission lines from ions like \ion{He}{i} formed at higher
temperatures are also red-shifted by several km\,s$^{-1}$ with respect to the
stellar velocity. Therefore these phenomena, indicating infall to regions at the
stellar surface that are offset relative to the rotational axis, appear to be a
universal property of active cTTS. 

\section{Discussion}
\label{sec:discussion}

\subsection{Modelling the RV variations}

Cross-correlation of DR Tau spectra with the synthetic spectrum in the region
6000--6040 \,\AA\, reveals that the bissector of cross-correlation
function (as described in Dall et al. \cite{dall06}) correlates with the
stellar radial velocity. This excludes the case of a binary motion and suggests
that RV variations are caused by distortions in the photospheric line profiles.
The same effect was found earlier for RU Lup by Stempels et al.
(\cite{stempels07}).

The anti-phase RV variations of the photospheric and emission lines suggest that
the line emission is confined to a certain area on the stellar surface, which is
not aligned with the axis of rotation, resulting in periodic shifts in RV as the
star rotates. The velocity curves in Fig.~\ref{rv_abs} have minima and maxima
separated by about 4.5 days, which is similar to the photometric and
spectroscopic periods (4.5--5.0 days) reported above, and which have been
linked to rotational modulation. 

The effect of an ``emission spot'' resulting in the anti-phase RV variations can
be illustrated with a simple model. We generated synthetic spectra of a single
rotating star plus an area emitting the narrow emission lines under the
assumption that this area has the form of a circular spot. The input parameters
for the model are: projected rotational velocity ($v\,\sin i$), inclination
($i$) of the stellar rotational axis to the line of sight, spot latitude
($\phi$), spot radius (r$_{\rm spot}$), local spectra of the photosphere and the
chromospheric spot, and the limb darkening coefficient ($\mu$). The spectrum was
integrated over the visible stellar surface at different phases of stellar
rotation. We then determined the radial velocities of the photospheric lines
RV$_{\rm phot}$ and the emission lines RV$_{\rm em}$ as a function of phase.

In this model most of the input parameters are restricted by observations. From
$v\,\sin i$ = 5 $\pm$ 1 km\,s$^{-1}$, rotational period 4.5--5.0 days and R$_*$
= 1.46 R$_{\sun}$, we obtain the inclination angle $i = 20\degr \pm 4\degr$, i.
e. the star is seen close to pole-on. The inclination is not very sensitive to
the adopted $A_V$, because the slopes of lines of equal radii in the $V$ versus
$(V-R)$ diagram and reddening are the same.

The spot latitude ($\phi$) on the stellar surface can be estimated from the 
amplitude $\Delta$RV of the radial velocity variations of the \ion{Fe}{i}
emission lines: $\Delta$RV$ = v \sin i\,\cdot\,\cos(\phi)$. With $v \sin i =
5.0 \pm 1.0$ km\,s$^{-1}$ and $\Delta$RV$ = 1.15 \pm 0.20$ km\,s$^{-1}$ we get
$\phi = 76\degr \pm 4\degr$.

A synthetic spectrum with $T_{\rm eff} = 4100$ K and $\log g = 3.7$ was taken as
the local spectrum of the undisturbed photosphere surrounding the spot. Within
the spot area the local spectrum was assumed to be a sum of the photospheric
spectrum and a line emision spectrum with similar line widths and line ratios.
The overall emission line strength was a free parameter. The limb darkening
coefficient was taken as $\mu$ = 0.8 (Al-Naimi~\cite{al-naimi78}).
Table~\ref{model_par} summarises the model parameters.

\begin{table}
\caption{Stellar and chromospheric spot model parameters }
\label{model_par}
\begin{center}
\begin{tabular}{c|c|c|c|c|c}
\hline
\hline
\multicolumn{6}{c}{\it Stellar parameters} \\
\hline
  $T_{\rm eff}$ & log $g$ & $L_*$   &  $R_*$   & $v\,\sin i$  &  $i$ \\
  (K)     &      & (L$_{\sun}$) &  (R$_{\sun}$) & (km\,s$^{-1}$) &  (deg) \\
\hline      
  4100   &  3.7   &  0.52   &  1.46     &  5.0     & 20 \\
\hline
\multicolumn{6}{c}{\it Chromospheric spot parameters} \\
\hline
$r_{\rm spot}$ & $\phi$ & \multicolumn{4}{c}{ } \\
(deg) & (deg) & \multicolumn{4}{c}{ } \\
\hline
14  & 76 & \multicolumn{4}{c}{ } \\
\hline
\end{tabular}
\end{center}
\end{table}

\begin{figure}
\centerline{\resizebox{9cm}{!}{\includegraphics{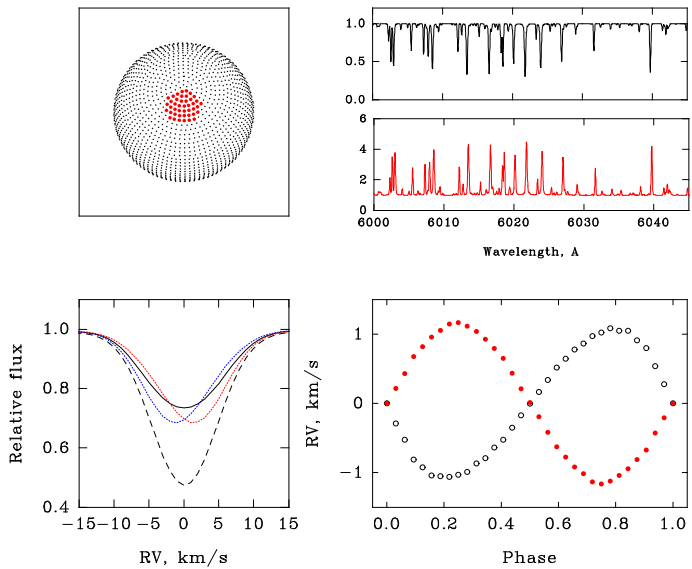}}}
\caption{Upper panels: location of the chromospheric spot and the local spectra
corresponding to the undisturbed photosphere (absorption) and to the spot
(emission). Lower panels: left -- profiles of the \ion{Ca}{i} $\lambda$ 6039
\,\AA\ line, corresponding to phases 0.0 spot facing the observer, solid line), 
0.22 and 0.78 (maximal blue- and red-shifted absorption, dotted lines). Dashed
profile: a normal photosphere without spot; right -- RV-variations of absorption
lines (open circles) and emission lines (filled circle).}
\label{dr_spmod}
\end{figure}

As shown in Fig.~\ref{dr_spmod} the model can reproduce the observed anti-phase 
RV-variations very well. The emission line strengths were tuned to obtain the
best fit to the observed RV amplitudes. Equal RV amplitudes of emission and
absorption lines occur when the latter is about halfway filled with emission
(i.e. VF$_{\rm line}$ $\approx$ 1). A spot radius of r$_{\rm spot}$ = 14$\degr$
(filling factor = 0.015) adopted in this model is about the same as the
typical radius of the accretion stream in the 3-D simulations of magnetospheric
accretion by Romanova et al. (\cite{romanova03}). The model also reproduces the
redshifted velocity offset of the He lines, provided that this emission is
formed in infalling gas, with a velocity vector directed towards the star.

In some other cTTS, the Zeeman-Doppler imaging technique has been used to
reconstruct the large-scale magnetic topology and to outline \ion{Ca}{ii} line
emission from the shocked gas near the stellar surface: V2129 Oph (Donati et
al.~\cite{donati07} and ~\cite{donati11}), BP Tau (Donati et
al.~\cite{donati08}), CV Cha and CR Cha (Hussain et al.~\cite{hussain09}), and
AA Tau (Donati et al.~\cite{donati10}). While the recovered structures are
sometimes complex, it is not uncommon to see hot spots, as indicated by the
enhanced \ion{Ca}{ii} emission. The adopted size and derived location relative
to the rotational axis for our \ion{Fe}{i} emitting area in DR Tau resembles
the spots of \ion{Ca}{ii} emission found in V2129 Oph and BP Tau.

It is worthwhile to note that the distortions of a photospheric line profile
caused by an emission core (Fig.~\ref{dr_spmod}) are very similar to those
caused by a dark spot. The two effects can be distinguished with the help of
photometry made simultaneously with spectropolarimetry. Although we cannot
discard a contribution from cool spots in the observed RV-variations of
photospheric lines in DR Tau, both the differential veiling and the anti-phase
RV-variations indicate that these are primarily caused by line emission. The
presence of narrow emission lines inside photospheric lines is particularly
evident in S CrA SE (Gahm et al.~\cite{gahm08}).

\subsection{Accretion powered emission} 

A direct signature of infall are the inverse P Cyg (IPC) absorption components.
In our series of spectra of DR Tau, IPC are present in the strongest 
\ion{Fe}{ii} lines (multiplet 42), in the \ion{Na}{i} D lines, in the higher
Balmer lines and, less evident, in the \ion{He}{i} line at 5876
\,\AA\, ({Fig.~\ref{dr_accr}}). In the low-veiling spectrum most emission lines
have faded out, and only the strongest emission lines are present. {\it On this
occasion the IPC profiles have also disappeared.} During this episode of
relatively weak accretion both the veiling continuum and the chromospheric
emission are considerably reduced. The upper limit to the veiling continuum
that day was only $\sim$ 0.5, as is evident from Fig.~\ref{vei_dep}. The area of
enhanced chromospheric emission in DR Tau is permanently present, but varies in
strength, and is probably coupled to a variable amount of mass accretion. 

We can roughly estimate the expected accretion luminosity for the low-veiling
spectrum using the relation between H$\alpha$ emission line flux and accretion
luminosity by Fang et al.~(\cite{fang09}). Our spectra are not flux-calibrated,
but for the low-veiling spectrum we can assume that DR Tau was close to its
minimum brightness (see Sect. 3.2). In the low-veiling spectrum, $EW(H\alpha) =
83.5$ \AA. For a K7V star with $L_*$ = 0.5 L$_{\sun}$, this EW corresponds to 
$log(L_{H\alpha}/L_{\sun}) = -2.53$. Using the relation by Fang et
al.~(\cite{fang09}), we obtain $log(L_{accr}/L_{\sun}) = -0.89 \pm 0.25$, or
$L_{accr}/L_* \approx 0.26$, which is compatible with an observed veiling $<0.5$
in the $V$ band.

\begin{figure}
\centerline{\resizebox{5cm}{!}{\includegraphics{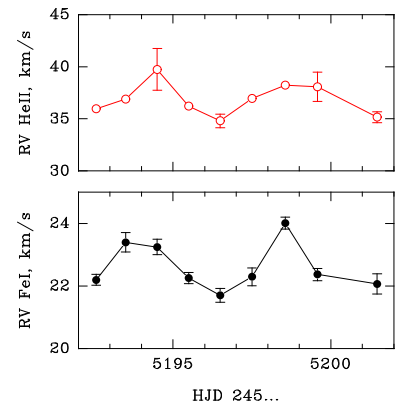}}}
\caption{Correlated variations of RV of the narrow emission of \ion{Fe}{i} and \ion{He}{ii}.}
\label{rv_he2}
\end{figure}

To explore further whether the narrow \ion{Fe}{i} emission is related to the hot
accretion spot, we followed radial velocity variations of the \ion{He}{ii}
$\lambda$ 4686\,\AA\, emission. In standard dipole magnetospheric accretion
models He is ionised by the X-ray radiation from the shock. Fig.~\ref{rv_he2}
shows that the RV of the narrow \ion{Fe}{i} emission lines vary {\it in phase}
with the narrow \ion{He}{ii} emission line. This strongly suggests that the area
of chromospheric emission is located close to the hot region at the footprint of
the accretion stream (see also discussions by Batalha et al. \cite{batalha96}
and Beristain et al. \cite{beristain98}). Moreover, the \ion{He}{ii} line is
clearly red-shifted relative to other narrow lines. As discussed by e.g. Lamzin
(\cite{lamzin98}), these ``high-temperature'' lines are formed in the post-shock
region. 

Our observations thus demonstrate that there is a close relation between the
appearance of enhanced chromospheric emission and accretion in classical T
Tauri stars. The narrow emission lines of e.g. \ion{Fe}{i} appears over an area
at the stellar surface that is associated with the footprints of the magnetic
funnels, and that is offset from the pole of rotation.

In standard dipole magnetospheric models the accretion spot may take the form of
a ring, which becomes distorted when the magnetic and rotational axes are not
aligned. However, recent models assume much more complex magnetic field
configurations (e.g. Mohanty \& Shu \cite{mohanty08}). The 3-D simulations of
magnetically channelled accretion by e.g. Romanova et al.
(\cite{romanova04,romanova11}) and Long et al. (\cite {long11}) reflect the
complex magnetic structures found from Doppler imaging of TTS. 

In the general concept of magnetically channeled accretion, the accretion spot
on a cTTS determines the observed brightness of the star, the veiling of
photospheric lines and the strength of certain emission lines, and all three
parameters must vary in correlation. However, observations do not always support
this prediction (see e.g. Petrov et al., \cite {petrov01}). In order to resolve
this discrepancy, we suggest that the area of enhanced chromospheric emission is
much more extended than the size of the hot accretion spot. In the models by
Romanova et al., less dense and non-uniform gas surrounds the main accretion
stream and falls onto a much larger area than what is covered by the main
stream, up to 16\% of the stellar surface. A possibility is that our observed
enhanced chromospheric emission is excited by such a milder accretion flow and
therefore distributed over a similar area surrounding the smaller hot spot. One
can assume that the accompanying shocks and related phenomena are less
pronounced than at the main footprints, and that little or no continuous excess
emission is released. In this scenario, a tight correlation between the
brightness, veiling and emission line strength is not expected.

X-ray observations of cTTS show that besides very powerful coronal emission
there is as a rule one component of soft X-rays that can be associated with the
hot spot (see e.g. G\"udel \& Telleschi \cite{gudel07a} and Argiroffi et al.
\cite{argiroffi09} and references therein). Recently, Brickhouse et al.
(\cite{brickhouse10}) presented X-ray spectroscopy of the cTTS TW Hya. A
low-temperature component (T $\sim$ 2.5 MK) could be identified with the
accretion shock, as depicted in the standard accretion models, and a
high-temperature component was ascribed to coronal emission at T $\sim$ 10 MK.
In addition, the authors discuss a third component, represented by low-density
lines of e.g. \ion{O}{vii} formed at T $\sim$ 1.75 MK, which fills a great
volume outside the post-shock region. Brickhouse et al. speculate that violent
mass outflows and shocks may propagate from the base of the accretion shock (see
also Orlando et al. \cite{orlando10}), and that the photospheric surroundings
are heated, which supplies ionised gas to accretion-fed magnetic structures,
such as magnetic loops. In addition, one can expect that the surface layers
surrounding the hot spot are excited by UV-light directly from the hot spot. 

Regardless of the dominating cause of the enhanced chromospheric emission one
would expect that the atmospheric structure adjacent to the hot spot is
affected. In some early attempts to explain line emission and veiling in cTTS,
Herbig (\cite{herbig70}) and Cram (\cite{cram79}, \cite{cram80}) introduced the
concept of a ''deep chromosphere'', where the temperature minimum occurs at
deeper layers in the atmosphere, thus resulting in the appearence of stronger
emisson spectrum. We speculate that our area of enhanced emission surrounding
the hot spot could be similar to a deep chromosphere with a temperature minimum
much deeper in the atmosphere than in a normal dwarf, but still above the
photospheric continuum. In any case, the chromospheric emission in cTTS appears
to be related not only to solar-like magnetic activity, but is to a considerable
extent powered by accretion processes. 

\section{Conclusions}
\label{sec:conclusions}

We have studied periodic changes in photospheric lines and narrow components of emission lines in DR Tauri and other classical T Tauri stars, and investigated the nature of the dilution of photospheric lines, 
the so-called veiling. In particular we note:

\begin{itemize}
\item  In the course of our observations the amount of veiling in DR Tau  varies
from practically nothing to factors of more than 10 times the stellar continuum
intensity. We have developed three methods to distinguish the contribution from
line emission and continuous excess emission to the veiling. We conclude that
narrow emission lines fill in the photospheric absorption lines, and that this
effect is more pronounced when the veiling is strong. Therefore, estimates of
mass accretion rates based on the assumption of a continuous veiling will be
overestimated for cTTS with strong emission line spectra.

\item Radial velocities of photospheric and narrow chromospheric lines vary in anti-phase in all cTTS studied here.  The mean velocity of the narrow lines are at rest relative to the stellar radial velocity. The velocity amplitudes are small. We have modelled these variations and infer that the emission lines are formed in an area at the stellar surface that is offset from the rotational axis. These properties appears to be a common phenomenon in cTTS.

\item The narrow  emission line of \ion{He}{ii} $\lambda$ 4686 \AA\ in DR Tau varies in phase with the narrow chromospheric components, but is systematically red-shifted by $\sim$ 10 km\,s$^{-1}$, which is
consistent with the origin of this line in decelerated post-shock gas. 
We conclude that the area of enhanced chromospheric emission is directly related to the hot spot area, and that chromospheric emission is induced by accretion. In accordance with this, 
we demonstrate that a sudden disappearance of inverse P Cygni profles in strong lines coincided 
with a fading of both the veiling and the strength of chromospheric lines.

\end{itemize}

We suggest that the area of enhanced chromospheric emission comes from an
extended region at the stellar surface surrounding the hot spot. The physical
mechanism that triggers the emission can be related to current ideas that differ
from the standard model in that accretion is more wide-spread, or emanates from
injection of mass and energy from the shocked region under the main accretion
stream. As a result the photosphere surrounding the hot spot is heated, which
leads to a modified atmospheric structure, 
which in turn produces the enhanced emission in narrow emission lines
responsible for the extra component in the veiling through line-filling of
photospheric absorption lines. 

\begin{acknowledgements}

This work was supported by the INTAS grant 03-51-6311. HCS acknowledges grant 621-2009-4153 of the Swedish Research Council.

\end{acknowledgements}

\end{document}